\documentclass[twocolumn,aps,epsf,showpacs,amssymb,amsmath,prl]{revtex4}
\usepackage{graphicx}
\usepackage{epsfig}

\newcommand{\be}{\begin{eqnarray}}
\newcommand{\ee}{\end{eqnarray}}

\newcommand{\maxflip}{
\includegraphics[width=0.4 cm]{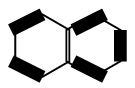}}
\newcommand{\maxfliptwo}{
\includegraphics[width=0.4 cm]{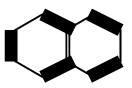}}
\newcommand{\doublehex}{\includegraphics[width=0.4 cm]{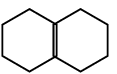}}

\begin{document}
\preprint{APS/123-QED}
\title{Variational wavefunction study of the triangular lattice supersolid}
\author{Arnab Sen$^1$, Prasenjit Dutt$^{1}$, Kedar Damle$^{1,2}$, and
R.~Moessner$^{3,4}$}

\affiliation{%
$^1$Department of Theoretical Physics,
Tata Institute of Fundamental Research, Mumbai 400005, India \\
$^2$ Physics Department, Indian Institute of Technology Bombay, Mumbai 400076, India\\
$^3$ Rudolf Peierls Centre for Theoretical Physics, Oxford University, Oxford OX1 3NP,
UK\\
$^4$ Max-Planck-Institut f\"ur Physik komplexer Systeme, 01189 Dresden, Germany
}%

\date{\today}

\begin{abstract}
We present a variational wavefunction which explains the behaviour of
the supersolid state formed by hard-core bosons on the triangular
lattice. The wavefunction is a linear superposition of {\em only and
all} configurations minimising the repulsion between the bosons (which it
thus implements as a hard constraint).
Its properties can be evaluated exactly---in particular, the
variational minimisation of the energy yields (i) the surprising and
initially controversial spontaneous density deviation from half-filling  (ii) a quantitatively accurate
estimate of the corresponding density wave (solid) order parameter.
\end{abstract}

\vskip2pc

\maketitle

There has been sustained interest in using
ultracold atoms confined in optical lattice potentials to realize strongly-correlated systems
of interest to condensed matter physics~\cite{RecentBECReview}.
The recently discussed possibility that Helium has a supersolid phase\cite{KITP}
leads, in this context, to a natural question: Can the
lattice analog of this, namely a superfluid
phase that simultaneously breaks lattice translation symmetry,
be seen in such optical lattice experiments?

Although other examples are known~\cite{KITP}, perhaps the best candidate for such a lattice supersolid persisting over a wide range of parameters (such as chemical potential and interaction energy) is
that observed numerically in several
recent Quantum Monte-Carlo (QMC) studies of a two-dimensional system of strongly interacting bosons in a triangular lattice potential~\cite{Wessel_Troyer,Heidarian_Damle,Melko_etal}, and confirmed
in subsequent follow-up work~\cite{Bonnisegni_Prokofiev}.
Following earlier work~\cite{Auerbach_Murthy}, these QMC studies considered the model Hamiltonian:
\be
H = -t\sum_{\langle ij \rangle}(b_{i}^{\dagger}b_{j} + b_{j}^{\dagger}b_i) + V\sum_{\langle ij \rangle } \left( n_i - 1/2 \right) \left(n_j - 1/2 \right)
\label{boson_H}
\ee
where $b_{i}^{\dagger} (b_i)$ is the boson creation (annihilation) operator, $t$ represents the strength of the boson hopping amplitude, $V$ is the strength of the nearest neighbour repulsion, and the bosons are restricted
to be in the hard-core limit ($n_i=0,1$) by a strong onsite repulsive term not written down explicitly (here, we have only displayed the Hamiltonian for the value of the chemical potential $\mu$ at which
the system has particle-hole symmetry).
Strikingly, in this particle-hole symmetric, hard-core case, an {\em extended} supersolid state  was observed numerically
for {\em all} $V/t \geq 8.9$, and seen to possess both a non-zero
superfluid stiffness, and density wave (`solid') order at the three sublattice
($\sqrt{3} \times \sqrt{3}$ ordering) wavevector $\vec{Q}$ (the state is stable to changes in $\mu$ as well).

\begin{figure}
 \includegraphics[width=\hsize]{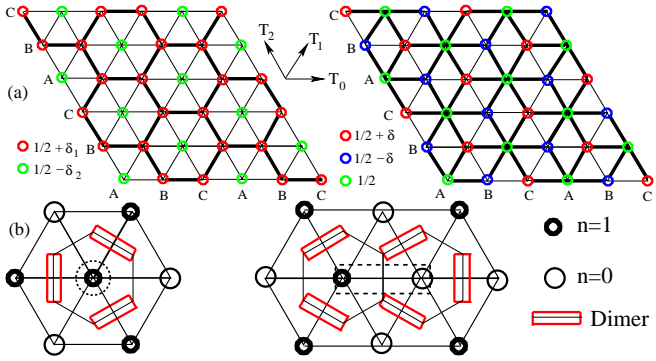}
      \caption{(color online). a) The two possible supersolid states at the
        three-sublattice wavevector $\vec{Q}$.
Dark (light) links represent higher (lower) bond kinetic energy, and sites
are color coded to reflect mean density. b)
      Flippable single and double hexagons in the dimer representation, and the corresponding local density configuration.}
      \label{thestates}
  \end{figure}

There are  two other genuinely surprisingly features of this phase.
The first~\cite{Heidarian_Damle} concerns the nature of the solid order. Plausible mean-field theory
arguments~\cite{Melko_etal} predict that the density wave order in the supersolid at $\mu=0$ should involve a `$(+ - 0)$' type three sub-lattice pattern of density with $\rho_a = 1/2$, $\rho_b=1/2+\delta$, $\rho_c=1/2-\delta$, while
in reality the system prefers a different `$(+ - -)$' pattern with the same ordering wavevector: $\rho_a=1/2 + \delta_1$, $\rho_b=\rho_c=1/2-\delta_2$, with $\delta_{1,2} > 0$ (Fig~\ref{thestates}).
General symmetry arguments~\cite{Melko_etal,Heidarian_Damle} predict that the {\em total} density of the system should also exhibit a spontaneous deviation from half-filling in a `$(+ - -)$' state. The second surprise is that although there is such a deviation, it is {\em extremely small}  in `natural' units (i.e compared to the strength of the density wave
order itself), and therefore extremely difficult to see numerically~\cite{Heidarian_Damle,Melko_etal}.

In order to understand this large $V/t$ supersolid phase in the hardcore limit, it is necessary to take into account this
strong repulsion as a hard constraint. Here we consider an analytically tractable variational wavefunction that takes into
account this hard constraint from the
start, and uses a variational parameter to tune the amplitudes of different minimum repulsion energy configurations.
The energetically optimal wavefunction accounts for both qualitative
(`$(+--)$' order) and quantitative (size of the order parameter) aspects
of the supersolid phase in the limit of strong nearest neighbour repulsion. Our treatment thus
represents a remarkable instance in which a non-trivial constraint {\em on the lattice scale} imposed by a dominant
term in the Hamiltonian can be taken into account in an exact manner in a variational calculation---for instance, the celebrated problem of
accounting for the single-occupancy constraint imposed (by strong coulomb repulsion) on holes in the $Cu$-$O$ planes of high-T$_c$ superconductors
has resisted analytical treatment thus far.

 Our starting point is the observation
that the supersolid ground state at large $V/t$ must have a wavefunction that lies entirely in the subspace
of minimum interaction energy configurations.
More precisely, in the classical ($t=0$) limit, the ground states are diagonal
in the particle-number basis, and form an extensively degenerate set of states corresponding
to all minimum repulsion  energy configurations of particles.
To leading order in $t/V$, the slow dynamics
induced by the kinetic energy term then leads to an effective Hamiltonian ${\cal H}$ acting in
within this manifold of states:
\be
{\cal H}_{eff} = -t \sum_{\langle ij \rangle} \mathcal{P}_g (b_{i}^{\dagger}b_{j} + b_{j}^{\dagger}b_{i}) \mathcal{P}_g
\label{eff_H}
\ee
where $\mathcal{P}_g$ is the projection operator to the minimum repulsion energy subspace.

As $H$ maps to a system of $S=1/2$ spins ($S^z_i = n_i -1/2$ where $n_i$
is the boson number at site $i$) with {\em frustrated}
antiferromagnetic exchange $J^z = V$ between the $z$ components of
neighbouring spins,
ferromagnetic exchange $J_\perp = 2t$ between their $x$ and $y$ components,
and magnetic field $B_z \propto \mu$, the ground states in this $t=0$ limit may be identified with minimally frustrated states of the classical Ising antiferromagnet on the triangular lattice~\cite{Wannier}.
As is well-known, these
may be conveniently characterized in terms of dimer coverings of the dual honeycomb lattice (with a dimer
placed on the dual link perpendicular to every frustrated bond of a given spin configuration).
In this language, ${\cal H}_{eff}$ is simply a quantum dimer model with a {\em ring-exchange term} that
acts on each pair of adjacent hexagons on the dual honeycomb lattice (corresponding to each bond
$\langle i j \rangle$ of the triangular lattice):
\be
{\cal H}_{eff} &=& -t \sum_{\doublehex_{\langle ij\rangle}} (|\maxflip_{\langle ij \rangle} \rangle \langle \maxfliptwo_{\langle ij \rangle} | + h.c.)
\label{dimer}
\ee

Thus, the supersolid behaviour at large $V/t$ should be understood in terms of the ground state
of this quantum dimer model. As was noted in earlier literature~\cite{Heidarian_Damle,Melko_etal},
a trial state obtained from an equal amplitude superposition of {\em all} minimum interaction energy
configurations (which maps, apart from a global particle-hole transformation,  to a uniform superposition of dimer covers of the honeycomb lattice) already
provides a substantial kinetic energy gain, while minimizing the inter-particle repulsion by construction.

Since $\langle \Psi_0| b_i^{\dagger}|\Psi_0\rangle$
is clearly proportional to the {\em non-zero} probability that hexagon $i$ is {\em
flippable} in the classical dimer model, this trial state immediately provides a rationale for
the persistence of off-diagonal long-range order and superfluidity in the large $V/t$ limit.
On the other hand, this trial state is unable to fully account for the density wave order of the true supersolid
state, as density correlators in $|\Psi_0\rangle$  map on to spin correlations
of the $T=0$ classical Ising antiferromagnet on the triangular lattice which does not support
genuine long-range order~\cite{Wannier}, but instead displays power-law order at
the three-sublattice wavevector~\cite{Stephenson}.\begin{figure}
\includegraphics[width=\hsize]{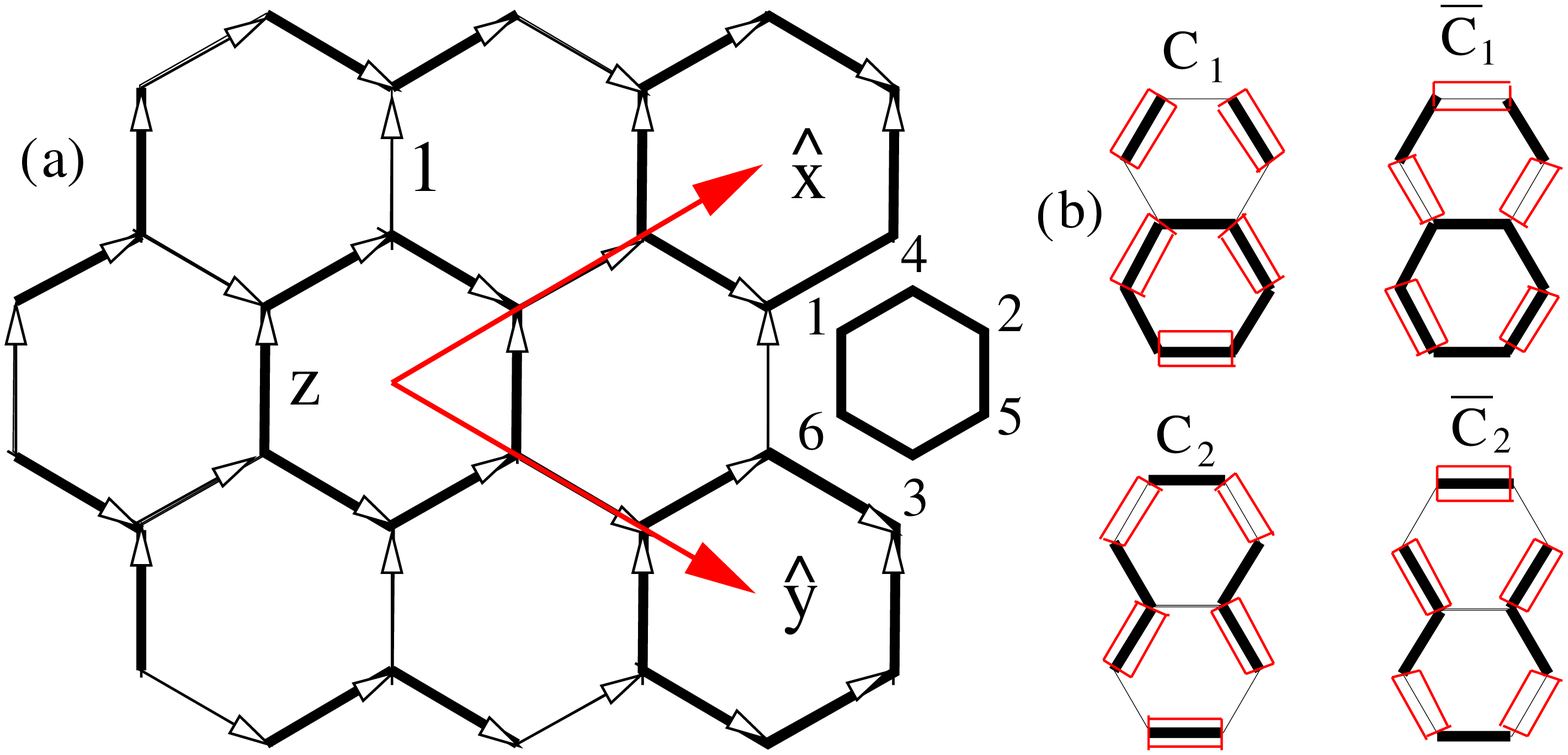}
      \caption{(color online). a) Dimer ensemble with a three-sublattice pattern of fugacities which breaks lattice translation
      symmetry at wavevector $\vec{Q}$.  b) The two types of flippable double-hexagons, with two flippable configurations each---note that light (dark) links correspond to fugacities $1$ ($z$).}
      \label{variationalstuff}
  \end{figure}

Here, we focus instead on a
variational state obtained from a dimer model with two types of links,
with different dimer fugacities $1$ and $z \ge 0$, defined on the
honeycomb lattice in a pattern (Fig \ref{variationalstuff}) which
breaks lattice symmetry at wavevector $\vec{Q}$ for all $z \neq 1$:
\be |\Psi_{var} (z)
\rangle = \sum_{C_d} \sqrt{{P_{C_d}(z)}/{2}} \left (| n_+(C_d)
\rangle + |n_-(C_d) \rangle \right )
\ee where $P_{C_d}(z)$ is the
probability of a dimer configuration $C_{d}$ in this dimer model, and
the two particle-hole conjugate configurations
$|n_\pm(C_d) \rangle$ corresponding to $C_d$
occur with the same amplitude in this nodeless wavefunction (this follows from the Perron-Frobenius theorem, as was noted earlier in a different context\cite{IQF}).

Clearly, $|\Psi(z)\rangle$ is characterized by three-sublattice
density wave order of the $(+ - -)$ type (see Fig~\ref{thestates}) for $z < 1$  while for $z>1$, it displays
density wave order of  the $(+ - 0)$ type.  Furthermore, as the state is constructed from
a coherent superposition of Fock states with a considerably wide distribution of average density, $|\Psi(z)\rangle$
is expected to also possess off-diagonal long range order associated with superfluidity, at least for
$z$ close to $z=1$.

In order to perform an unbiased variational study, we need to locate minima of the variational estimate of
energy per site (in units of $2t$): $E(z) = \langle  \Psi_{var} (z) | {\cal H}_{eff}|\Psi_{var} (z) \rangle/2tL^2$.
To calculate $E(z)$, we note that $ \langle  \Psi_{var} (z) | {\cal P}_g b^{\dagger}_i b_j {\cal P}_g |\Psi_{var} (z) \rangle
= \sqrt{P(\maxflip_{\langle i j \rangle}) P(\maxfliptwo_{\langle i j \rangle})}$, where $P(\maxflip_{\langle i j \rangle})$
and $P(\maxfliptwo_{\langle i j \rangle})$ are the probabilities that the double hexagon $\langle i j \rangle$ is
in the {\em flippable} configurations depicted in their respective arguments. This allows us to write
\be
E(z) & = & -\left(2\sqrt{P(1{\cal C})P(1{\cal \bar{C}})} + \sqrt{P(2{\cal C}) P(2{\cal \bar{C}})} \right )
\ee
where $P(1{\cal C})$ and $P(1{\cal \bar{C}})$ ($P(2{\cal C})$ and $P(2{\cal \bar{C}})$) are the respective probabilities that type $1$ (type $2$) double-hexagons
are in flippable configurations ${\cal C}$ and ${\cal \bar{C}}$ (see Fig~\ref{variationalstuff}).
Thus, the variational energy
is lowered if the expected number of {\em flippable double-hexagons} is large.
Since the number of flippable double hexagons must decrease rapidly to zero in the $z \rightarrow 0$ as
well as the $z \rightarrow \infty$ limit as the dimers freeze into a perfect columnar or plaquette state in
these limits, it is  immediately clear that one or more variational minima of $E(z)$ must occur at or in the vicinity
of the translationally invariant point $z=1$.

 Calculation of the probabilities $P$ is greatly facilitated by the well-known formulation of
dimer models on planar graphs in terms of Grassmann variables~\cite{Kastelyn,Samuel,Moessner_Sondhi}: For the
case at hand, this can be obtained by using the `arrow convention' displayed
in Fig~\ref{variationalstuff} to define an antisymmetric matrix $M$, with
$M_{ij} = + \mu_{\langle ij \rangle}$ (where the fugacity associated with link $\langle i j \rangle$)  if an arrow points from point $i$ to $j$ and $M_{ij} = -\mu_{\langle ij \rangle}$ if the arrow goes from $j$ to $i$ ($M_{ij} = 0$ if $i$,$j$ are not nearest neighbours,
$\mu_{\langle i j \rangle}$ equals $z$ or $1$ as shown in Fig~\ref{variationalstuff}).
The dimer partition function $Z$ is then obtained as
$Z =\left |Pf[M] \right | = \left | \int [\mathcal{D}\psi] \exp(S) \right |$, where $Pf$ denotes the Pfaffian and
$S = \sum_{i<j}M_{ij}\psi_i \psi_j$ is the action for Grassmann variables $\psi_i$ defined on sites of the honeycomb lattice.\begin{figure}
\includegraphics[width=\hsize]{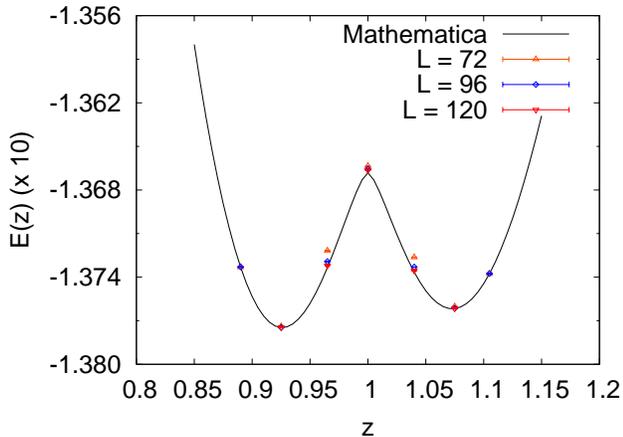}
      \caption{(color online). Variational energy per site in units of $2t$ as a function of $z$.}
      \label{energy}
  \end{figure}

To proceed further, we represent the honeycomb net as an underlying triangular Bravais
lattice with Roman letter coordinates (representing centers of those hexagons whose links all
have fugacity $z$) that is decorated by six basis points (corresponding to vertices of such hexagons) labeled by Greek
letters (Fig~\ref{variationalstuff} ), and write the action as $S = \frac{1}{2} \sum_{\vec{x}, \alpha} \sum_{\vec{y}, \beta}M^{\alpha,\beta}_{\vec{x},\vec{y}} \psi_{\alpha, \vec{x}} \psi_{\beta, \vec{y}}$.
Transforming to Fourier space,
we obtain
$S = \frac{1}{2}\sum_{\vec{k},\alpha,\beta} \tilde{M}^{\alpha,\beta}_{\vec{k}}\psi_{\alpha,\vec{k}}\psi_{\beta, -\vec{k}}$,
with the $6 \times 6$ matrix $\tilde{M}^{\alpha,\beta}_{\vec{k}}$ is given as
\begin{displaymath}
\mathbf{\tilde{M}}(\vec{k}) =
\left( \begin{array}{cc}
{\mathbf{ 0}} & {\mathbf{ \tilde{R}}(\vec{k})} \\
{\mathbf{ -\tilde{R}^{\dagger}}(\vec{k})} & {\mathbf{ 0}}
\end{array} \right)
\end{displaymath}
where ${\mathbf{ 0}}$ is the $3 \times 3$ null matrix and ${\mathbf{\tilde{R}}}$ can be written as
\begin{displaymath}
\mathbf{\tilde{R}}(\vec{k}) =
\left( \begin{array}{ccc}
z & -e^{-ik_y} & -z \\
 -z & -z & e^{ik_x} \\
 -e^{-ik_x+ik_y} & z & -z
\end{array} \right)
\end{displaymath}
\begin{figure}
\includegraphics[width=\hsize]{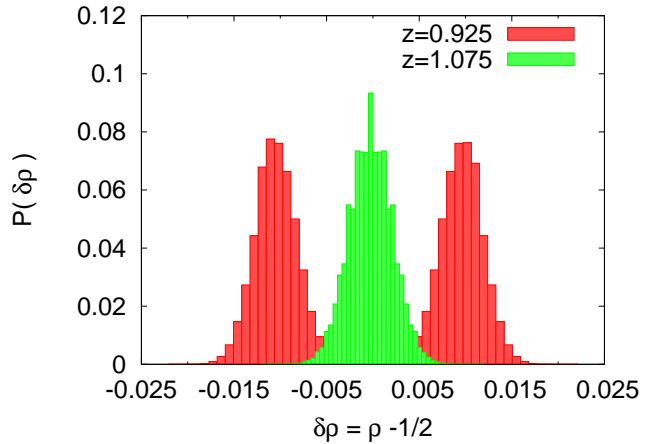}
      \caption{(color online). Distribution of the total density at the global
        variational minimum at $z \approx 0.925$, as well as at the competing
      local minimum at $z \approx 1.07$. Data shown is for $L=96$.}
      \label{ferrimagnet}
  \end{figure}

Next, we note that this Grassmann formulation provides a simple prescription for the
probability that five dimers occupy alternate links $l_1 \dots l_5$ on the perimeter of double hexagon
made up of points $i=1,2\dots 10$:
\be
P(\{ l_1 \dots l_5\}) = \left(\prod_{m=1}^{5}\mu_{l_m} \right)\cdot |\langle \psi_1 \psi_2 \psi_3 \cdot \cdot \psi_{10} \rangle|
\label{answer}
\ee
where $\mu_{l_m}$ are the fugacities of the $5$ links occupied by dimers.
This allows us to write $E(z)$ as
\be
E(z)  =   -2z^{\frac{7}{2}} |\langle \psi_1 \psi_2 \psi_3 \cdot \cdot \psi_{10} \rangle_{1}| - z^{3} |\langle \psi_1 \psi_2 \psi_3 \cdot \cdot \psi_{10} \rangle_{2}|
\label{finalanswer}
\ee
where
the subscripts on the $10$-point correlators refer to the type of double-hexagon on which they are calculated~(see Fig~\ref{variationalstuff}).

The $10$ point correlators involved in this expression can be computed in the free field action $S$
using Wick's theorem and knowledge of the two point correlators:
\be
\langle \psi_{\alpha,\vec{x}}\psi_{\beta,\vec{y}}\rangle = \int \frac{d^2k}{4\pi^2} \exp(-i\vec{k} \cdot (\vec{x} - \vec{y}))  (-{\mathbf M}^{-1}(-\vec{k}))_{\alpha,\beta}
\label{}
\ee
where the integration is over the Brillouin zone $k_x,k_y \epsilon (-\pi,\pi]$.
As $\langle \psi_i \psi_j \rangle = 0$ if $i$ and $j$ belong to the same sublattice on the honeycomb lattice, one needs to keep track of `only'  $5! = 120$ contractions in evaluating either of the
two $10$-point correlators. Keeping track of these contractions and performing the required
$k$ integrations using MATHEMATICA, we obtain the variational energy $E(z)$ shown
in Fig~\ref{energy}. As is clear from this figure, $E(z)$ has two minima, a global minimum at $z \approx 0.9250$, and local minimum at $z \approx 1.0750$ with energy only very slightly higher (by about $\approx 0.047 \%$)
than its value at the global minimum.

Thus, our variational calculation yields a $(+ - -)$ type three sublattice ordered supersolid state in the large $V/t$
limit, and suggests that the energy of the $(+ - 0)$ supersolid at the same wavevector is only
slightly higher than that of the  $(+ - -)$ supersolid. In order to obtain an independent verification of
this result, we have also used the method of Ref~\cite{Sandvik_Moessner} to simulate the corresponding
dimer model and numerically calculate $E(z)$---the results of these calculations are seen to match precisely with
the results obtained by Grassmann techniques (Fig~\ref{energy}).

This simulation also allows us
to measure directly the solid order parameter $\psi = m_a + m_b e^{2\pi i/3} + m_c e^{4\pi i/3}$, where $m_a,m_b,m_c$ are the number densities on the three sublattices of the triangular lattice.
We obtain $|\psi|^2 = 0.03885$ for the global minimum at $z = 0.9250$, which agrees within 10\% with the results from QMC \cite{Melko_etal}
extrapolated to $V/t \rightarrow \infty$.
( the value in the competing $(+-0)$ state is $|\psi|^2 = 0.03697$ at the slightly higher energy local minimum
at $z = 1.0750$). In addition, we also measure the histogram of the total density, from which
one may calculate the ground state expectation values of all powers of the density in the supersolid state.
While the local minimum at slightly higher energy has no spontaneous deviation of
density from half-filling,
we find that the density histogram at the global minimum shows a characteristic two-peak structure,
reflecting a very small ($\sim 2\%$) spontaneous deviation from half-filling characteristic of the $(+--)$
supersolid (Fig~\ref{ferrimagnet}).

To obtain further insight into the smallness of this spontaneous density deviation from half-filling, we note that
reducing $z$ slightly from $z=1$ introduces a field coupled to the Fourier component of the particle-density at wavevector
$\vec{Q}$. The effect of this field can be understood by using the  well-known~\cite{Nienhuis_Hillhorst_Blote} height model formulation of $z=1$ dimer model in terms of an action $S_h = \kappa \int d^2 x (\nabla h)^2$ for a `height' field
$h(x)$ in terms of which one may write the Fourier component of the particle density at wavevector $\vec{Q}$
as $\rho_{\vec{Q}} \sim exp(i\pi h/3)$, and at wavevector $0$
as $\rho_{tot} \sim exp(i \pi h)$.

For $z < 1$ but close to $z=1$, the response of
the density at these two wavevectors can be obtained by calculating the corresponding susceptibilities,
and one finds that the response at wavevector $\vec{Q}$ is much larger than
at wavevector $0$, thereby providing a rationalization for the smallness of this symmetry allowed
density deviation (in comparison with the size of the order parameter).

Thus, our variational approach provides a coherent picture of  the triangular lattice supersolid at large $V/t$: Kinetic energy
effects are seen to select a $(+--)$ supersolid state with a three-sublattice density wave
order parameter comparable in magnitude to the mean density itself. In addition, the competing $(+-0)$ supersolid
is seen to be very close in energy to this variational ground state, consistent with numerical evidence
and analytical arguments~\cite{Burkov_Balents} that the order parameter phase (which distinguishes
between these two states)
is only weakly pinned in the supersolid state.
Furthermore, the spontaneous
deviation of the density from $1/2$ is seen to be a very small fraction ($\sim 2\%$) of
the mean density, consistent with the surprisingly small value obtained in earlier
QMC studies.

{\em Acknowledgements: } We thank D.~Dhar for very insightful discussions
and S. L. Sondhi for collaboration on closely related work. We
acknowledge computational resources at TIFR and Oxford and
support from DST SR/S2/RJN-25/2006 (KD), the John Fell OUP Research
Fund--Oxford-India network in Theoretical Physical Sciences (AS), and
a British Council  RXP grant (RM).

\end{document}